\documentstyle[preprint,epsfig,aps]{revtex}

\begin{document} 


\title{Non-collapsing renormalized QRPA with proton-neutron pairing \\
for neutrinoless double beta decay
\footnote{Supported by the ``Deutsche 
Forschungsgemeinschaft'', contract No. Fa 67/17-1 and
by the ``Graduierten Kolleg - Struktur und Wechselwirkung 
von Hadronen und Kernen'', DFG, Mu 705/3.}}

\author{F. \v Simkovic$^1$
\thanks{{\it On leave from:}Bogoliubov Theoretical Laboratory, 
Joint Institute for Nuclear Research, 
141980 Dubna, Moscow Region, Russia and Department of Nuclear Physics,  
Comenius University, Mlynsk\'a dolina F1, Bratislava, Slovakia}, 
J. Schwieger$^1$, M. Veselsk\'y$^2$, 
G. Pantis$^3$ and Amand Faessler$^1$ }
\address{1.  Institute f\"ur Theoretische Physik der Universit\"at 
T\"ubingen\\ 
Auf der Morgenstelle 14, D-72076 T\"ubingen, Germany \\
2. Department of Nuclear Physics, Comenius University, \\
Mlynsk\'a dolina F1, Bratislava, Slovakia \\
3.  Theoretical Physics Section, University of Ioannina,\\
GR 451 10, Ioannina, Greece }
\date{\today}
\maketitle
\begin{abstract}
Using the renormalized quasiparticle random phase approximation
(RQRPA), we calculate the light neutrino mass mediated mode of
neutrinoless double beta decay ($0\nu\beta\beta$-decay) of $^{76}Ge$,
$^{100}Mo$, $^{128}Te$ and $^{130}Te$. Our results indicate that the 
simple quasiboson approximation is not good enough to study the
$0\nu\beta\beta$-decay, because its solutions collapse for physical
values of $g_{pp}$. We find that extension of the Hilbert space
and inclusion of the Pauli Principle in the QRPA
with proton-neutron pairing, allows us to extend our calculations 
beyond the point of collapse, for physical values of
the nuclear force strength. As a consequence one might be able to
extract more accurate values on the
effective neutrino mass by using the best available
experimental limits on the half-life of $0\nu\beta\beta$-decay.
\end{abstract}

Neutrinoless double beta decay ($0\nu\beta\beta$), which involves the
emission of two electrons and no neutrinos, requires the neutrino to
be a Majorana particle with non-zero mass. This process violates the
lepton number conservation and occurs in some theories beyond the
standard model (see e.g. \cite{1}-\cite{4} for reviews).  The
$0\nu\beta\beta$-decay is not yet observed. The experimental
lower limits on the half-lives of some nuclei which undergo double
beta decay, provide us with the most stringent limits on the effective
mass of the electron neutrinos and the parameters of the right-handed currents
after evaluating the corresponding nuclear matrix elements.

The calculation of the $0\nu\beta\beta$-decay matrix elements is based
on some approximation schemes, from which the Quasiparticle Random
Phase Approximation (QRPA) is the most prominent one
\cite{5}-\cite{8}, despite an uncertainty in the calculation of the
nuclear matrix element, which is related to its sensitivity on the
renormalization of the particle-particle component of the residual
interaction. Thus in the framework of
proton-neutron QRPA (pn-QRPA) using the zero range delta-force Engel
et al. \cite{6} have found a strong suppression of the
$0\nu\beta\beta$-decay matrix elements by introducing short-range
correlations, whereas Tomoda et al. \cite{5} and Muto et al. \cite{7}
by using a more realistic effective NN interaction arrived at the
opposite conclusion.

Recently new approaches for the calculation of the nuclear many body
Green function have been proposed, which are supposed to offer more
reliable results than that of pn-QRPA. The pn-QRPA has been extended
to include proton-neutron pairing (full-QRPA)
\cite{9}.  It has been found that proton-neutron pairing
influences the $0\nu\beta\beta$-decay rates significantly \cite{10}.
Toivanen and Suhonen \cite{11}, Schwieger, \v Simkovic and Faessler
\cite{12} and F. Krmpoti\'c et al. \cite{13} have studied the effect of the 
violation of the Pauli principle in the correlated ground state 
of the pn-QRPA introducing the renormalized pn-QRPA (pn-RQRPA). 
Schwieger et al. \cite{12} have done calculations 
of the two-neutrino double beta decay matrix elements
using this renormalized pn-QRPA but also including 
proton-neutron pairing (full-RQRPA). 
They have shown \cite{12} that the inclusion of both p-n pairing and 
an improved treatment of the Pauli
principle removes the difficulties arising from the strong dependence
of the matrix element
on the particle-particle strength $g_{pp}$ in the standard pn-QRPA.

The $0\nu\beta\beta$-decay allows to determine an upper limit of the
effective Majorana electron neutrino mass $<m_{\nu}>$ \cite{1}-\cite{4}. 
For this process the most stringent experimental lower limit of the 
life-time
for the $0\nu\beta\beta$-decay of $^{76}Ge$ favours especially
this nucleus for extracting an upper limit of
$<m_{\nu}>$. The purpose of this
article is to study in detail the effects of the Pauli principle, the
proton-neutron pairing and the truncation of the nuclear
Hamiltonian in the evaluation of the nuclear matrix element governing
the $0\nu\beta\beta$-decay of $^{76}Ge$. In addition the full-RQRPA
will also be used to calculate $0\nu\beta\beta$-decay of $^{100}Mo$,
$^{128}Te$ and $^{130}Te$. The stability of the nuclear matrix elements in
respect to the renormalization of the particle-particle force will be
discussed.

The full-RQRPA, which describes the excited states of the even-even
nucleus, has been studied in Ref. \cite{12}. Therefore, here we
shall present only the formulae relevant to this work.

In the notation of Ref. \cite{12}, the half life of the
$0\nu\beta\beta$-decay in the case of the neutrino mass mechanism is
given by,
\begin{equation}
[T_{1/2}^{0\nu}]^{-1} = G_{01} (M^{0\nu}_{mass})^2 
\left(\frac{<m_\nu>}{m_e}\right)^2 
\label{eq:6}  
\end{equation}
with the effective neutrino mass
 \begin{equation}
<m_\nu> = \sum_{{j}} |U^{}_{{ej}}|^2 
m_{{j}} e^{ {{i}}\alpha_{{j}} }, 
\label{eq:7}   
\end{equation}
where $\exp({i}\alpha_{{j}})$ is the CP eigenvalue of the neutrino
mass eigenstate $|\nu_{{j}}>$, $U_{ei}$ is the element of the unitary
neutrino mixing matrix and $m_e$ is the mass of the electron.
$G_{01}$ is the integrated kinematical factor for the
$0^+_i\rightarrow 0^+_f$ transition \cite{1}-\cite{4,10}.  
The nuclear matrix element
$M^{0\nu}_{mass}$ consists of Fermi and Gamow-Teller contributions
\begin{equation}
M^{0\nu}_{mass}=M^{0\nu}_{GT}-\left(\frac{g_V}{g_A}\right)^2M_F^{0\nu},
\label{eq:8}   
\end{equation}
where
\begin{equation}
 \left .
 \begin{array}{c}
  M_F^{0\nu} \\
  M^{0\nu}_{GT}
 \end{array}
 \right\}
= \left\langle H(r_{12}) 
\begin{array}{c}
1 \\
{\bf \sigma}_1 \cdot {\bf\sigma}_2
\end{array}
\right\rangle,
\label{eq:9}   
\end{equation}
\begin{eqnarray}
<O_{12}>&=&
\sum_{{k l \acute{k} \acute{l} } \atop {J^{\pi}
m_i m_f {\cal J}  }}
~~(-)^{j_{l}+j_{k'}+J+{\cal J}}(2{\cal J}+1)
\left\{
\begin{array}{ccc}
j_k &j_l &J\\
j_{l'}&j_{k'}&{\cal J}
\end{array}
\right\}\nonumber \\
&&~~~~~~~~~~~
\times<pk,pk';{\cal J}|f(r_{12})\tau_1^+ \tau_2^+ {\cal O}_{12}
f(r_{12})|nl,nl';{\cal J}>
\nonumber \\
&&\times < 0_f^+ \parallel 
\widetilde{[c^+_{pk'}{\tilde{c}}_{nl'}]_J} \parallel J^\pi m_f>
<J^\pi m_f|J^\pi m_i>
<J^\pi m_i \parallel [c^+_{pk}{\tilde{c}}_{nl}]_J \parallel 
0^+_i >.
\label{eq:10}   
\end{eqnarray}
The short-range correlations
between the two interacting nucleons are taken into account by
a correlation function $f(r_{12})$ \cite{5,8}.
The neutrino-potential $H(r_{12})$ in the case of light
neutrino exchange takes the form
\begin{equation}
H(r)=\frac{2}{\pi}
\frac{R}{r} \int_{0}^{\infty} 
\frac{\sin(qr)}{q+(\Omega^{m_i}_{J^\pi}+\Omega^{m_f}_{J^\pi})/2}
\frac{1}{(1+q^2/{\Lambda}^2)^4}
d{q}.
\label{eq:11}   
\end{equation}
Here, $R=r_0A^{1/3}$ is the nuclear radius ($r_0=1.1 $ fm) and the
parameter $\Lambda $ of the dipole shape nucleon form factor is
chosen to be 0.85 GeV \cite{8,10}.  
In the full-RQRPA for the matrix elements of the one-body transition
densities we obtain 
\begin{eqnarray}
<J^\pi m_i \parallel [c^+_{pk}{\tilde{c}}_{nl}]_J \parallel 0^+_i>&=&
\sqrt{2J+1}\sum_{\mu,\nu = 1 , 2} m(\mu k,\nu l) 
\left [
u_{k \mu p}^{(i)} v_{l \nu n}^{(i)} {\overline{X}}^{m_i}_{\mu\nu}(k,l,J^\pi)
\right.\nonumber \\ && \left.
+v_{k \mu p}^{(i)} u_{l \nu n}^{(i)} {\overline{Y}}^{m_i}_{\mu\nu}(k,l,J^\pi)
\right ]
\sqrt{{\cal D}^{(i)}_{k \mu l \nu l J^\pi}},
\label{eq:13}   
\end{eqnarray}
\begin{eqnarray}
<0_f^+ \parallel \widetilde{ [c^+_{pk'}{\tilde{c}}_{nl'}]_J} 
\parallel J^\pi m_f>&=&
\sqrt{2J+1}\sum_{\mu,\nu=1,2}m(\mu k',\nu l') 
\left [
v_{k' \mu p}^{(f)} u_{l' \nu n}^{(f)} 
{\overline{X}}^{m_f}_{\mu\nu}(k',l',J^\pi)
\right.\nonumber \\ && \left.
+u_{k' \mu p}^{(f)} v_{l' \nu n}^{(f)} 
{\overline{Y}}^{m_f}_{\mu\nu}(k',l',J^\pi)
\right ]
\sqrt{{\cal D}^{(f)}_{k' \mu l' \nu J^\pi}},
\label{eq:14}   
\end{eqnarray}
with 
$m(\mu a,\nu b)=\frac{1+(-1)^J\delta_{\mu \nu}\delta_{ab}}{(1+\delta_{\mu\nu}
\delta_{ab})^{1/2}}$. We note that the  
${\overline{X}}^{m}_{\mu\nu}(k,l,J^\pi)$ and
${\overline{Y}}^{m}_{\mu\nu}(k,l,J^\pi)$ amplitudes are calculated 
by the renormalized QRPA equation only for the configurations 
$\mu a \leq \nu b$ ( i.e., 
$\mu = \nu$ and the orbitals are ordered $a \leq b$ and 
$\mu = 1$, $\nu = 2$ and the orbitals are not ordered) \cite{12}.
For different configurations 
${\overline{X}}^{m}_{\mu\nu}(k,l,J^\pi)$ and
${\overline{Y}}^{m}_{\mu\nu}(k,l,J^\pi)$ in Eqs. 
(\ref{eq:13}) and (\ref{eq:14}) are given by following the prescription
in Eqs. (65) and (66) of Ref. \cite{8}.
The index i (f) indicates that the quasiparticles and the excited
states of the nucleus are defined with respect to the initial (final)
nuclear ground state $|0^+_i>$ ($|0^+_f>$). The overlap between two
intermediate nuclear states belonging to two different sets is given
in Ref. \cite{12}.

We note that for ${\cal D}_{k \mu l \nu J^\pi}=1$ the expressions
(\ref{eq:13}) and (\ref{eq:14}) are just the one body transition
densities in the full-QRPA \cite{10}. If in addition $u_{\text{2p}} =
\upsilon_{\text{2p}} = u_{\text{1n}} = \upsilon_{\text{1n}} = 0$ (i.e.
there is no proton-neutron pairing), Eqs.\ (\ref{eq:13}) and
(\ref{eq:14}) reduce to the expressions of the pn-QRPA \cite{5,7}.

We apply the pn-QRPA, full-QRPA (with proton-neutron pairing), 
pn-RQRPA and full-RQRPA methods to
the $0\nu\beta\beta$-decay of $^{76}Ge$ to study the effects of the
proton-neutron pairing and the Pauli-principle violation on the
nuclear matrix element $M^{0\nu}_{mass}$. In previous calculations 
that model space has been used,
which described best the beta strength
distributions for the initial and final nuclei. However, there is a
substantial difference between the beta strength matrix element and
the double beta decay matrix element. Unlike  the case of the double beta
decay, which is a second order process, meson exchange currents
do not play an important role in the calculation of the single beta strength
distribution.  The meson exchange currents are incorporated in our 
 calculation
through the G-matrix elements of the nuclear Hamiltonian. We suppose that
they can be taken into account in a proper way only if the inclusion
of Pauli principle is considered, and a large enough model space is
assumed.  In addition we wish to stay as closely as possible to the 
Brueckner reaction matrix of the Bonn potential. A smaller model space
requires a more drastic renormalization of the force. Since one does not
know a reliable microscopic theory for the renormalization as a 
function of the model space, a smaller model space means more
free parameters.
Thus we enlarge the model space of the nuclear Hamiltonian.
The model spaces used are the following:\\
$\underline{9-\text{level model space}}:$ The full $3\hbar\omega$ and
$4\hbar\omega$ major oscillator shells. 
This model space has been used
by Tomoda et al. \cite{5} and by Muto et al. \cite{7}.\\
$\underline{12-\text{level model space}}:$ It consists of the full
$2-4\hbar\omega$ major oscillator shells.\\ 
$\underline{21-\text{level
model space}}:$ This model space extends over the $0-5\hbar\omega$
major oscillator shells.

The single particle energies have been calculated with a
Coulomb-corrected Woods-Saxon potential. As a realistic two-body
interaction we use the nuclear matter G-matrix calculated from the
Bonn one-boson-exchange potential \cite{14}. 
We introduce the
parameterization $d_{\alpha}G$ ($\alpha$=pp, nn and pn), $g_{ph}G$ and
$g_{pp}G$ for the pairing, particle-hole and particle-particle
interaction, respectively. 
The quasiparticle energies and amplitudes have been found by solving the
HFB-equation with proton-neutron pairing\cite{9}. 
For a model space larger than two oscillator major shells we neglect
the mixing of different "n" but the same "ljm" orbitals. We suppose
that the Woods-Saxon potential is already a good potential and therefore 
shell mixing is not appreciably affecting the double beta decay.
The renormalization of the pairing
interaction has been determined to 
fit the empirical proton, neutron and proton-
neutron pairing gaps according to Ref. \cite{9}. 
The renormalization parameter $d_{\alpha}$ together with the
experimental proton ( $\Delta ^{exp}_{p}$ ), neutron 
( $\Delta ^{exp}_{n}$ ) and proton - neutron ( $\delta
^{exp}_{pn}$ ) pairing gaps for $^{76}Ge$ and $^{76}Se$ are listed
in Table \ref{table1}. We see that $d_{pp}$ and $d_{nn}$ values are
close to unity. A $d_{pn}$ value higher than unity is the price
paid for the spherical symmetry of the model which excludes the 
treatment of the T=0 pairing. The J=0 T=0 pairs can be treated
in a BCS or even HFB approach only due to deformation. The T=0 
pairing is effectively taken into account by the renormalization 
of the T=1 J=0 n-p interaction leading to a higher value of $d_{pn}$. 
We note that in the framework of the HFB method 
we have found that the importance of the proton-neutron pairing for the
ground state properties of nuclei is decreasing significantly as isospin
increases \cite{9}. This fact is in agreement with the recent
sophisticated study of J. Engel et al. \cite{15}. 

Further, we fixed
$g_{ph}=0.8$ as in our previous calculations \cite{10} and only the
particle-particle interaction strength $g_{pp}$ is considered as a
variable. Fig. 1 (a) shows the calculated nuclear matrix element
$M^{0\nu}_{mass}$ in the framework of the pn-QRPA for the
$0\nu\beta\beta$ of $^{76}Ge$ as function of $g_{pp}$. It is
worthwhile to notice that by increasing the model space
$M^{0\nu}_{mass}$ becomes extremely sensitive to the $g_{pp}$ and even
within the physically acceptable region of $g_{pp}$ collapses and
crosses zero. This behaviour has its origin in the contribution of the
$J^\pi=1^+$ intermediate states of the odd-odd nucleus to the many
body Green function $M^{0\nu}_{mass}$, which becomes too large because
of the generation of too many ground state correlations.  This feature
has not been found in the previous calculations as the model spaces
used there, were too small. We surmise that the different suppressions
of the $0\nu\beta\beta $-decay matrix elements found by Tomoda et al.
\cite{5} and Muto et al. \cite{7} on one side and Engel et al.
\cite{6} on the other side could have their origin not only in the
different interactions but also in the different model spaces used.
Unfortunately, the model space of Ref. \cite{6} is not specified to
draw a definite conclusion. F. Krmpoti\'c and S. Sharma 
\cite{16} have calculated $0\nu\beta\beta$-decay of $^{76}Ge$ by  
using eleven dimensional model space (The full $3\hbar\omega$ and
$4\hbar\omega$ major oscillator shells plus $0h_{9/2}$ and 
$0h_{9/2}$ orbitals.), zero range $\delta$-force interaction 
and different treatment of the two-nucleon correlation function. 
Their results show also a strong suppression of the $M^{0\nu}_{mass}$
with increasing strength of the particle-particle force. 
To our knowledge, we are the first to study in details the
extension of the Hilbert space for the $0\nu\beta\beta$-decay of
$^{76}Ge$ and showing that its solution collapse for physical
value of $g_{pp}$ in the QRPA. 
Therefore, pn-QRPA method does not allow us to make definite
$0\nu\beta\beta $-decay rate predictions.

In Fig. 1 (b) we show the results obtained in the framework of the
full-QRPA, which take into account proton-neutron pairing \cite{10}.
The $M^{0\nu}_{mass}$ has been calculated till the collapse of the
full-QRPA, which occurs for the $J^\pi=0^+$ or $2^+$ channel. The
early collapse of the full-QRPA for the 21-level model space does not
allow us to draw conclusions inside the expected physical range $0.8
\le g_{pp} \le 1.2$ within this approximation.

The nuclear matrix element $M^{0\nu}_{mass}$ obtained within the
pn-RQRPA is shown in Fig. 1 (c). From the comparison with the Fig. 1
(a) it follows that the inclusion of ground state correlations beyond
the QRPA in the calculation of $M^{0\nu}_{mass}$ removes the
difficulties associated with the strong dependence of
$M^{0\nu}_{mass}$ on the particle-particle strength. It means that the
Pauli principle plays a prominent role in the evaluation of the
nuclear matrix elements governing the second order processes. We
conclude, that the rather big difference between the results of
9-level and 12-level model spaces and the small difference between the
results of 12-level and 21-level model spaces favours the 12-level
model space for the study of double beta decay of $^{76}Ge$.

In Fig. 1 (d) we present the results with the
full-RQRPA method.  This method takes into account both proton-neutron
pairing and the Pauli exclusion principle and does not show a
breakdown for any realistic interaction strength.  A comparison with
Fig. 1 (c) shows that proton-neutron pairing decreases the values of
$M^{0\nu}_{mass}$ for the small model space (9-levels) and increases
them for bigger, more realistic model spaces. Thus the full-RQRPA  
allows an extension of the calculations beyond the point of collapse 
and supports more stable solutions for calculating the
 $M^{0\nu}_{mass}$ ,
 with respect to the renormalization of the
particle-particle force.

In order to gain more confidence in the full-RQRPA method, we have also
performed calculations for $0\nu\beta\beta$-decay of $^{100}Mo$,
$^{128}Te$ and $^{130}Te$. As noticed already in Ref. \cite{10},
the evaluation of $M^{0\nu}_{mass}$ is very sensitive to the truncation
of the model space in the case of $^{100}Mo$. We have found that the
model spaces used in the previous calculations were not sufficient to
stabilize the results and therefore we have performed calculations
with the 21-level model space defined above. For the Te isotopes we
have introduced a  model space, which comprises 20 levels: The full
2-5 $\hbar\omega$ major oscillator shells and $0i_{11/2,13/2}$
subshells. The calculated matrix elements $M^{0\nu}_{mass}$ are shown
in Fig. 2. For all nuclei studied  $M^{0\nu}_{mass}$ is only
weakly dependent on the  strength of particle-particle
interaction. This allows us to have more confidence in the limits for the
effective electron neutrino mass $<m_\nu>$extracted by using the best presently
available experimental limits on the half-lives of
the $0\nu\beta\beta$-decays.  We obtain the following limits:
\begin{eqnarray}
^{76}Ge: ~~~ <m_\nu>~<~1.3~eV~~~~[T^{0\nu-exp}_{1/2}~>~5.6\times10^{24}
~\text{Ref. [17]}],\nonumber \\
^{100}Mo: ~~~ <m_\nu>~<~2.4~eV~~~~[T^{0\nu-exp}_{1/2}~>~4.4\times10^{22}
~\text{Ref. [18]}],\nonumber \\
^{128}Te: ~~~ <m_\nu>~<~1.1~eV~~~~[T^{0\nu-exp}_{1/2}~>~7.3\times10^{24}
~\text{Ref. [19]}],\nonumber \\
^{130}Te: ~~~ <m_\nu>~<~4.5~eV~~~~[T^{0\nu-exp}_{1/2}~>~2.3\times10^{22}
~\text{Ref. [20]}].\nonumber
\end{eqnarray}
We see, that by increasing significantly the model space for
$^{100}Mo$ the value of $M^{0\nu}_{mass}$ is much bigger as in our
previous calculations \cite{10}. The rather big energy release for the
$0\nu\beta\beta$-decay of $^{100}Mo$ and the large value of the
matrix element 
$M^{0\nu}_{mass}$ favour this nucleus for further experimental study.

The above results suggest, that in the calculation of the
nuclear matrix element $M^{0\nu}_{mass}$, it is necessary to take
into account the Pauli principle and to consider a large enough model
space in addition to pn-pairing.  As it was noticed already by J.
Hirsch, P. Hess and O. Civitarese \cite{21} and by F. Krmpoti\'c, A.
Mariano, E.J.V. Passos, A.F.R. de Toledo Piza and T.T.S. Kuo
 \cite{13}, the price which is paid for taking into
account the ground state correlation beyond the QRPA, is the violation
of the Ikeda Sum Rule. In the framework of the full-RQRPA we have
\begin{eqnarray}
S^- -S^+  ~&=& ~ \sum_{k\mu l\nu} |<k\parallel \sigma \parallel l>|^2 
~D_{k\mu l\nu 1^+}~ \nonumber \\
&&\times (v^2_{k\mu n}-v^2_{k\mu p}
+v_{k\mu p}u_{k\mu n}v_{l\nu p}u_{l\nu n}-
u_{k\mu p}v_{k\mu n}u_{l\nu p}v_{l\nu n}),
\end{eqnarray}
which yields 3(N-Z) only when $D_{k\mu l\nu 1^+}=1$ (quasiboson
approximation).  If the usual HFB constraint on the particle number is
used and if the model space contains both states $j_{+1/2, -1/2}=l\pm
1/2$ of the spin-orbit splitting, the QRPA
fulfills the Ikeda Sum Rule independent of the chosen model space,
which is not a realistic feature of the model. In the RQRPA we
necessarily face its violation because of $D_{k\mu l\nu 1^+}\neq 1$. By
using the full-RQRPA for $^{76}Ge$ and $^{76}Se$ we have found $S^-
-S^+ $ to be $0.8-0.9\times 3(N-Z)$ in the physical acceptable region
of the parameter $g_{pp}$. It is supposed that the omission of the
scattering terms in the operators for $\beta^+$ and $\beta^-$
transitions, could be the reason for this small mismatch
\cite{13,21}. It is worthwhile mentioning that 
the Ikeda Sum Rule is not satisfied even if the BSC equations
are solved within the RQRPA with the condition that the average 
particle number in the correlated ground state is conserved
\cite{13}.  
Nevertheless we expect that a small violation of the Ikeda Sum 
Rule wont alter our main results.

In summary, we have studied the effects of Pauli principle violation
and proton-neutron pairing on the calculation of the
$0\nu\beta\beta$-decay matrix element. We have shown that the 
renormalised QRPA allows extending the calculations beyond the collapse 
and thus that the simple
quasiboson approximation is not sufficiently accurate to calculate
reliable the nuclear many-body Green function governing the
$0\nu\beta\beta$-decay process. This becomes evident, if a larger, more
realistic, Hilbert space is considered.  We have found that the
$0\nu\beta\beta$-decay matrix element calculated via the full-RQRPA,
which includes the Pauli effect of fermion pairs, avoids the collapse
for physical values of the nuclear force, and therefore we can argue
that we have some evidence that the Pauli effect violation inherent
in the traditional QRPA affects the   $0\nu\beta\beta$-decay 
matrix element significantly.
As a consequence, to our opinion, the results presented in the text	
for the effective neutrino mass for $g_{pp}=1$ are more accurate. 

Note added. After completing this work, we recieved a preprints by 
J. Hirsch et al. \cite{22} and by J. Engel et al. \cite{23} who studied the
validity of the renormalized QRPA within a schematic exactly solvable
models, which are not intended to reproduce actual nuclear properties. 
They confirmed that the renormalized QRPA offers advantages over the
QRPA. However, they
founded some discrepancies between the exact and the RQRPA solutions
after the point of collapse of the QRPA. It is questioned 
whether these discrepancies should hold in more realistic calculations. 
If they would, it just supports our conclusion that we need a theory 
in which the Pauli principle is explicitly incorporated. In the 
renormalized QRPA the Pauli principle is considered in an 
approximate way. We note that there is no exactly solvable realistic
model.

\newpage

\widetext
\begin{table}[t]
\caption{Experimental proton ( $\Delta ^{exp}_{p}$ ), neutron 
( $\Delta ^{exp}_{n}$ ) and proton - neutron ( $\delta
^{exp}_{pn}$ ) pairing gaps and
renormalization constants of the proton - proton (
$d_{pp}$ ),
neutron - neutron ( $d_{nn}$ ) and proton - neutron ( $d_{pn}$ )
pairing interactions for studied nuclei.}
\label{table1}
\begin{tabular}{ccccc} 
Nucleus ~ ( ${\Delta }^{exp}_{p}$, ${\Delta }^{exp}_{n}$, ${\delta }
^{exp}_{pn}$ ) & model & $d_{pp}$ & $d_{nn}$ & $d_{pn}$ \\ 
~~~~~~~ ( [MeV] ) & space & & & \\ \tableline
 $ ^{76}_{32}Ge_{44}$ ~ (1.561, 1.535, 0.388) 
 & 9 level & 1.063 & 1.238 & 2.093 \\ 
 & 12 level & 0.988 & 1.150 & 1.777 \\
 & 21 level & 0.899 & 1.028 & 1.506 \\ 
 $ ^{76}_{34}Se_{42}$ ~ (1.751, 1.710, 0.459) 
 & 9 level & 1.135 & 1.255 & 1.678 \\ 
 & 12 level & 1.027 & 1.182 & 1.524 \\
 & 21 level & 0.934 & 1.059 & 1.325 \\ 
 $ ^{100}_{42}Mo_{58}$ ~ (1.612, 1.358, 0.635) 
 & 21 level & 0.980 & 0.923 & 1.766 \\
 $ ^{100}_{44}Ru_{56}$ ~ (1.548, 1.296, 0.277) 
 & 21 level & 1.002 & 0.945 & 1.568 \\
 $ ^{128}_{52}Te_{76}$ ~ (1.127, 1.177, 0.149) 
 & 20 level & 0.873 & 0.942 & 1.780 \\
 $ ^{130}_{52}Te_{78}$ ~ (1.043, 1.180, 0.090) 
 & 20 level & 0.835 & 0.945 & 1.87 \\
 $ ^{128}_{54}Xe_{74}$ ~ (1.307, 1.266, 0.199) 
 & 20 level & 0.920 & 0.972 & 1.530 \\
 $ ^{130}_{54}Xe_{76}$ ~ (1.299, 1.243, 0.190) 
 & 20 level & 0.914 & 0.974 & 1.614 \\
\end{tabular}
\end{table}

\newpage

\begin{figure}[htb]
\caption{The calculated nuclear matrix element $M^{0\nu}_{mass}$
for the $0\nu\beta\beta$-decay of $^{76}Ge$ as a function of the
particle-particle interaction strength $g_{pp}$. In (a), (b), (c) and
(d) $M^{0\nu}_{mass}$ have been calculated within the pn-QRPA,
full-QRPA, pn-RQRPA and full-RQRPA, respectively. The dotted line
corresponds to the 9-level model space, the dashed line to the
12-level model space and the solid line to 21-level model space,
respectively.}
\label{figa}
\end{figure}

\begin{figure}[htb]
\caption{The calculated nuclear matrix element $M^{0\nu}_{mass}$ 
for the $0\nu\beta\beta$-decay of $^{76}Ge$ within the full-RQRPA as
function of particle-particle interaction constant $g_{pp}$.  In the
case of $^{76}Ge$ and $^{100}Mo$ ($^{128}Te$ and $^{130}Te$) the
21-level (20-level) model space has been used.  }
\label{figb}
\end{figure}

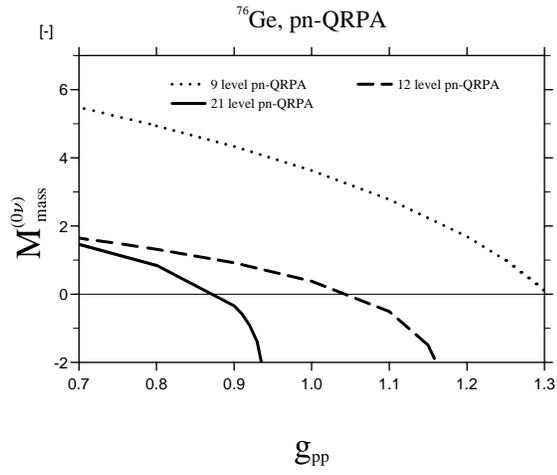
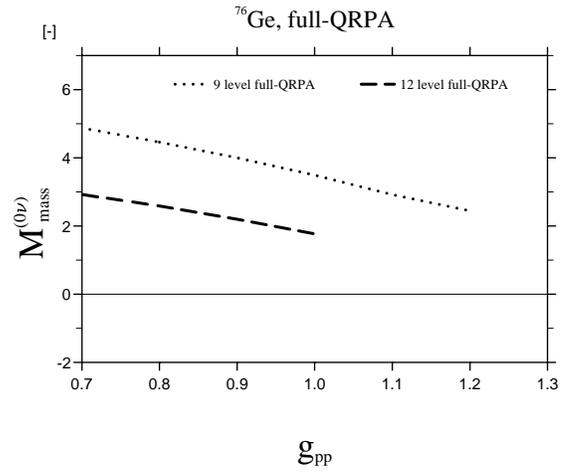
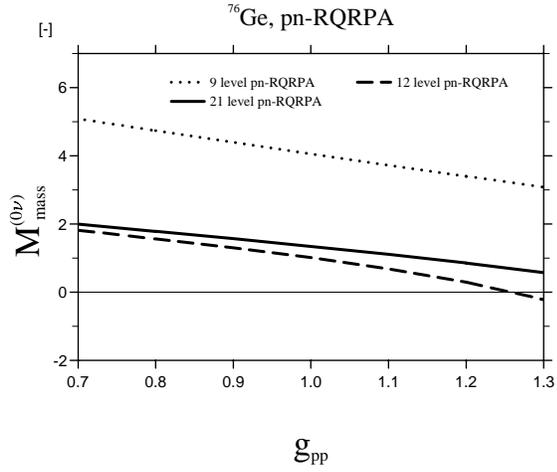
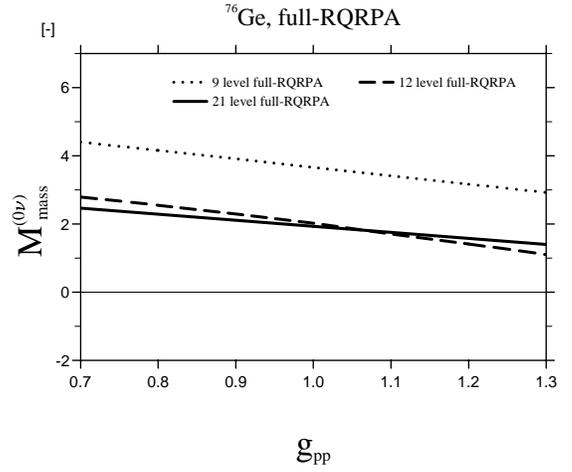
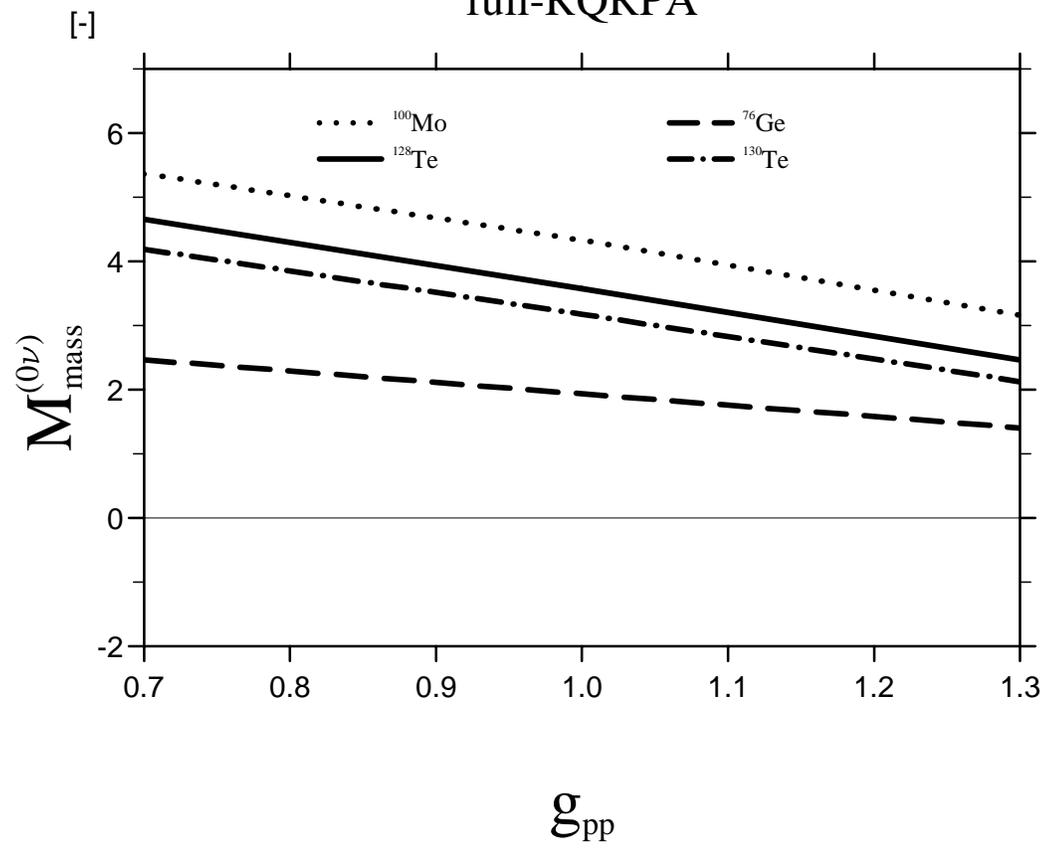
\end{document}